\documentclass[useAMS,usenatbib]{mn2e-1}
\usepackage{graphicx}
\usepackage{array,longtable}
\usepackage{natbib}
\usepackage{float}
\usepackage[T1]{fontenc}
\usepackage{ae}
\usepackage{aecompl}
\usepackage{amsmath, amssymb, amsfonts}
\usepackage{multicol}

\def\aj{AJ}%
\def\apj{ApJ}%
\def\apjl{ApJ}%
\def\apjs{ApJS}%
\def\aap{A\&A}%
\def\mnras{MNRAS}%
\def\pasp{PASP}%
\def\zap{ZAp}%
\def\nat{Nature}%
\title[Red Giant Branch Bump Star Counts]{Red Giant Branch Bump Star Counts in Data and Stellar Models}
\author[Nataf]{David M. Nataf$^1$\thanks{Email: david.nataf@anu.edu.au}
\vspace*{6pt}\\
$^{1}$Research School of Astronomy and Astrophysics, The Australian National University, Canberra, ACT 2611, Australia  }
\begin{document}
\include{journaldefs}
\date{Accepted 2014 September 19.  Received 2014 August 23; in original form 2014 July 16}

\pagerange{\pageref{firstpage}--\pageref{lastpage}} \pubyear{2014}
\maketitle
\label{firstpage}

\begin{abstract}
We compare model predictions to observations of star counts in the red giant branch bump (RGBB) relative to the number density of first-ascent red giant branch at the magnitude of the RGBB, $EW_{RGBB}$. The predictions are shown to exceed the data by $(5.2 \pm 4.3)$\% for the BaSTI models and by  $(17.1 \pm 4.3)$\% for the Dartmouth models, where the listed errors are purely statistical.  These two offsets are brought to zero if the Galactic globular cluster metallicity scale is assumed to be overestimated by a linear shift of $\sim 0.11$ dex and $\sim 0.36$ dex respectively. This inference based on RGBB star counts goes in the opposite direction to the increase in metallicities of ${\Delta}$[M/H]$\approx$0.20 dex that would be required to fix the offset between predicted and observed RGBB luminosities. This comparison is a constraint on ``deep-mixing" models of stellar interiors, which predict decreased RGBB star counts. We tabulate the predictions for RGBB star counts as a function of [Fe/H], [$\alpha$/Fe], CNONa, initial helium abundance, and age. Though our study suggests a small zero-point calibration issue, RGBB star counts should nonetheless be an actionable parameter with which to constrain stellar populations in the differential sense. The most significant outliers are toward the clusters NGC 5024 (M53), NGC 6723, and NGC 7089 (M2), each of which shows a $\sim 2 \sigma$ deficit in their RGBB star counts.
\end{abstract}
\maketitle


\section{Introduction}
\label{sec:introduction}
The red giant branch bump (RGBB) is a feature of colour-magnitude diagrams (CMDs) of old ($t \gtrsim 1$ Gyr), well-populated stellar populations. As a star first ascends the red giant branch (RGB), the location of hydrogen burning in the star, a shell of mass (0.001-0.0001) M$_{\odot}$, moves outward with time as the star expands and grows more luminous. When the hydrogen-burning shell approaches the discontinuity in the hydrogen abundance near the maximum depth reached by the convective envelope, the luminosity of the star temporarily drops before increasing again, leading to an excess in the luminosity function over an underlying exponential distribution in magnitudes. This excess is referred to as the RGBB \citep{1990ApJ...364..527S,1997MNRAS.285..593C,2006ApJ...641.1102B}.  The RGBB was first  theoretically described by \citet{1967ZA.....67..420T} and \citet{1968Natur.220..143I}. It was first empirically confirmed nearly two decades later,  by \citet{1985ApJ...299..674K}, in their observations of the globular cluster (GC) 47 Tuc. It has since been well-documented that the characteristic luminosity \citep{1990A&A...238...95F,1997MNRAS.285..593C,2003A&A...410..553R,2006ApJ...641.1102B,2010ApJ...712..527D,2011A&A...527A..59C,2011PASP..123..879T}, normalisation \citep{2001ApJ...546L.109B,2003A&A...410..553R,2006ApJ...641.1102B,2011ApJ...730..118N}, and shape \citep{2002ApJ...565.1231C} of this excess are a steeply-sensitive function of stellar models as well as assumed population parameters such as metallicity, age, and helium abundance.  

There has been significant progress within the literature in tracking how the luminosity of the RGBB depends on metallicity, and by now a clear picture has emerged.  When the RGBB is investigated in the Galactic GC population, the luminosity is found to steeply decline with metallicity \citep{2002A&A...391..945P,2010ApJ...712..527D,2011PASP..123..879T,2011A&A...527A..59C}, at a rate of dM$_{V,\rm{RGBB}}$/d[M/H] $=(0.737\pm 0.024)$ mag dex$^{-1}$ \citep{2013ApJ...766...77N} for a sample of Galactic GCs observed with the WFPC2 \citep{2002A&A...391..945P} and ACS  \citep{2007AJ....133.1658S,2011ApJ...738...74D} Galactic GC treasury programs. A similar trend is also observed for nearby dwarf galaxies \citep{2010ApJ...718..707M}. 

However, though the existence of the RGBB and its approximate trend of luminosity with metallicity is a spectacular confirmation of stellar theory, there is one caveat: stellar models seemingly overpredict the luminosity of the RGBB in both Galactic GCs and dwarf spheroidal galaxies. The typical discrepancy measured is $\sim$0.20 mag, increasing to $\sim$0.40 mag for metallicities [M/H]$=-$2.00 and below. This has been confirmed whether one compares the brightness of the RGBB to that of the ZAHB \citep{2010ApJ...712..527D}, the main-sequence turnoff \citep{2011A&A...527A..59C}, or the brightness of the point on the main-sequence at the same colour as the RGBB \citep{2011PASP..123..879T}. The combination of these three tests with their distinct systematics demonstrates that the discrepancy is not due to factors such as mass-loss on the RGBB, incorrectly-assumed ages, initial helium abundances, distances, or reddening estimates. Other possibilities, such as a large calibration error in the Galactic GC metallicity scale, or non-local overshoot at the base of the outer convective envelope \citep{1991A&A...244...95A,2012ApJ...746...20K} remain as plausible solutions, though the problem is not conclusively solved at this time. 

This overprediction of the luminosity of the RGBB is thus a symptom of currently unspecified problems in either stellar models or how we interpret them (e.g. the assumed metallicity scale), and additional tests would thus be beneficial. An alternative to comparing luminosity measurements and predictions is to do the same for the normalisation of the RGBB, i.e. the star counts, this would probe related physics but in principle have different sensitivies. Whereas the luminosity of the RGBB accounts for the \textit{location} of the abundance discontinuity within the star, the normalisation of the RGBB will be proportional to the \textit{amplitude} of that discontinuity \citep{2001ApJ...546L.109B,2006ApJ...641.1102B}, as the decreased molecular weight will slow down the efficiency of nuclear burning. 

The comparison between RGBB star counts in models and data has previously been looked at \citep{2001ApJ...546L.109B,2002A&A...391..945P,2006ApJ...641.1102B}, where agreement between theory and observation was found. However, there has since been an evolution in the Galactic GC metallicity scale \citep{2009A&A...508..695C}. Further, there has been a vast improvement in both the quality and the quantity of the available photometric data \citep{2007AJ....133.1658S,2010ApJ...708..698D} as well as in the methodology employed to measure the RGBB \citep{2013ApJ...766...77N}. The combination of these two factors should lead to a $\sim$300\% increase in total effective precision. There is thus sufficient incentive to revisit the issue. 

In addition to these factors, the Galactic GC system from which the bulk of the data for RGBB studies derive has been shown in the past decade to be more complex than previously believed. Many -- perhaps all -- GCs have a diverse star-formation history, with some GCs showing spectacular variations in CNO and initial helium abundances, such as $\omega$ Centauri \citep{2004ApJ...612L..25N} and NGC 2808 \citep{2007ApJ...661L..53P}. This may allow further testing of RGBB physics as continuously-improving CMDs of these GCs better separate the distinct RGBs, and thus allow measurements of two (or more) distinct RGBBs. Indeed, \citet{2011ApJ...736...94N} reported a gradient in $EW_{RGBB}$ and $I_{RGBB}$ toward 47 Tuc consistent with a helium-enhanced second generation being more centrally concentrated, i.e. the RGBB toward the inner part of 47 Tuc appeared slightly brighter and less well-populated. Though the difference in helium between the first and second generation of 47 Tuc are not large (${\Delta}Y \sim 0.02$, see \citealt{2012ApJ...744...58M}), the cluster is well-populated and has very low differential reddening, allowing precise measurements of relatively small changes.

In this investigation we tabulate predicted values of RGBB star counts by means of its equivalent width on the RG luminosity function, $EW_{\rm{RGBB}}$, from the BaSTI isochrones \citep{2004ApJ...612..168P}. We compare these to the values in a sample of 44 Galactic GCs where the RGBB's properties were measured by \citet{2013ApJ...766...77N}. 

\section{Data}
\label{sec:Data}
\begin{figure}
\begin{center}
\includegraphics[totalheight=0.17\textheight]{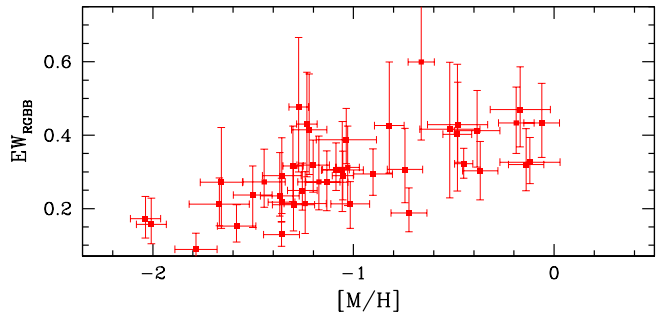}
\end{center}
\Large
\caption{\large  The trend of the equivalent width of the RGBB, $EW_{\rm{RGBB}}$, versus metallicity. Empirical values (red squares)  and corresponding error bars are from \citet{2013ApJ...766...77N}. } 
\label{Fig:EWRGBbumpVsMetDATA}
\end{figure}

The data used by this investigation is that of 44 ``gold-sample" measurements of $EW_{\rm{RGBB}}$ with 1-$\sigma$ errors from \citet{2013ApJ...766...77N}, where the methodology is to be found in Section 3 of that investigation. Following that paper, RGBB parameters from the GCs NGC 2808, 5286, 6388, and 6441 are excluded from the analysis due to the confounding issue of potentially very distinct multiple stellar populations within each of those clusters. These measurements were made on data from the \textit{Hubble Space Telescope} (HST) treasury programs of \citet{2002A&A...391..945P}, \citet{2007AJ....133.1658S}, and \citet{2011ApJ...738...74D}. We list the data in Table \ref{table:EWRGBBempirical} and we plot its scatter in Figure \ref{Fig:EWRGBbumpVsMetDATA}. 

\begin{table*}
\caption{\large Summary of empirical values used in this work. Maximum-likelihood values of the equivalent width of the RGBB, $EW_{\rm{RGBB}}$, along with 1-$\sigma$ upper and lower bounds are taken from the investigation of \citet{2013ApJ...766...77N}, which is based on \textit{Hubble Space Telescope} photometry from \citet{2002A&A...391..945P}, \citet{2007AJ....133.1658S}, and \citet{2011ApJ...738...74D}.   Globular cluster metallicities are taken from  \citet{2009A&A...508..695C},  except for that of Lynga 7 which is taken from \citet{2008A&A...479..741B}. } 
\large
\centering 
\begin{tabular}{lccccc} 
\hline\hline\hline 
Cluster Name & [M/H] & $\sigma_{\rm{[M/H]}}$   & EW$_{\rm{RGBB}}$ & EW$_{\rm{RGBB}}^{+} $ & EW$_{\rm{RGBB}}^{-}$  \\
 \hline\hline
LYNGA07 & $-0.38$ &  $0.11$ & 0.412 & 0.522 & 0.313 \\ 
NGC 104 (47 Tuc) & $-0.45$ &  $0.05$ & 0.322 & 0.364 & 0.282 \\ 
NGC 362 & $-1.09$ &  $0.07$ & 0.306 & 0.379 & 0.249 \\ 
NGC 1261 & $-1.02$ &  $0.10$ & 0.213 & 0.273 & 0.146 \\ 
NGC 1851 & $-0.90$ &  $0.10$ & 0.294 & 0.363 & 0.236 \\ 
NGC 3201 & $-1.27$ &  $0.05$ & 0.477 & 0.666 & 0.300 \\ 
NGC 5024 (M53) & $-1.79$ &  $0.11$ & 0.089 & 0.133 & 0.053 \\ 
NGC 5272 (M3) & $-1.26$ &  $0.07$ & 0.249 & 0.324 & 0.196 \\ 
NGC 5634 & $-1.66$ &  $0.11$ & 0.272 & 0.421 & 0.146 \\ 
NGC 5824 & $-1.67$ &  $0.15$ & 0.212 & 0.284 & 0.153 \\ 
NGC 5904 (M5) & $-1.05$ &  $0.05$ & 0.290 & 0.357 & 0.224 \\ 
NGC 5927 & $-0.06$ &  $0.09$ & 0.433 & 0.541 & 0.340 \\ 
NGC 5986 & $-1.37$ &  $0.10$ & 0.235 & 0.353 & 0.179 \\ 
NGC 6093 (M80) & $-1.58$ &  $0.10$ & 0.152 & 0.211 & 0.109 \\ 
NGC 6139 & $-1.44$ &  $0.11$ & 0.273 & 0.362 & 0.204 \\ 
NGC 6171 (M107) & $-0.66$ &  $0.07$ & 0.599 & 0.883 & 0.387 \\ 
NGC 6205 (M13) & $-1.36$ &  $0.07$ & 0.218 & 0.283 & 0.165 \\ 
NGC 6218 (M12) & $-1.03$ &  $0.06$ & 0.312 & 0.425 & 0.208 \\ 
NGC 6229 & $-1.17$ &  $0.11$ & 0.273 & 0.398 & 0.197 \\ 
NGC 6254 (M10) & $-1.30$ &  $0.05$ & 0.316 & 0.425 & 0.219 \\ 
NGC 6284 & $-1.06$ &  $0.10$ & 0.305 & 0.437 & 0.192 \\ 
NGC 6304 & $-0.14$ &  $0.09$ & 0.320 & 0.418 & 0.249 \\ 
NGC 6341 (M92) & $-2.01$ &  $0.07$ & 0.157 & 0.228 & 0.105 \\ 
NGC 6352 & $-0.48$ &  $0.07$ & 0.403 & 0.593 & 0.248 \\ 
NGC 6356 & $-0.12$ &  $0.15$ & 0.327 & 0.394 & 0.268 \\ 
NGC 6362 & $-0.82$ &  $0.07$ & 0.426 & 0.600 & 0.297 \\ 
NGC 6402 (M14) & $-1.13$ &  $0.11$ & 0.272 & 0.357 & 0.195 \\ 
NGC 6541 & $-1.50$ &  $0.10$ & 0.237 & 0.316 & 0.148 \\ 
NGC 6569 & $-0.48$ &  $0.15$ & 0.429 & 0.544 & 0.326 \\ 
NGC 6584 & $-1.24$ &  $0.11$ & 0.214 & 0.301 & 0.132 \\ 
NGC 6624 & $-0.19$ &  $0.09$ & 0.433 & 0.531 & 0.350 \\ 
NGC 6637 (M69) & $-0.37$ &  $0.09$ & 0.303 & 0.383 & 0.223 \\ 
NGC 6638 & $-0.74$ &  $0.09$ & 0.307 & 0.419 & 0.216 \\ 
NGC 6652 & $-0.52$ &  $0.15$ & 0.416 & 0.598 & 0.229 \\ 
NGC 6681 (M70) & $-1.36$ &  $0.10$ & 0.290 & 0.393 & 0.187 \\ 
NGC 6723 & $-0.72$ &  $0.09$ & 0.188 & 0.256 & 0.137 \\ 
NGC 6752 & $-1.23$ &  $0.05$ & 0.430 & 0.571 & 0.286 \\ 
NGC 6760 & $-0.17$ &  $0.15$ & 0.470 & 0.586 & 0.368 \\ 
NGC 6864 (M75) & $-1.04$ &  $0.15$ & 0.387 & 0.473 & 0.323 \\ 
NGC 6934 & $-1.30$ &  $0.11$ & 0.211 & 0.311 & 0.139 \\ 
NGC 6981 & $-1.22$ &  $0.09$ & 0.415 & 0.566 & 0.293 \\ 
NGC 7006 & $-1.20$ &  $0.08$ & 0.318 & 0.387 & 0.245 \\ 
NGC 7078 (M15) & $-2.04$ &  $0.08$ & 0.173 & 0.233 & 0.120 \\ 
NGC 7089 (M2) & $-1.36$ &  $0.09$ & 0.129 & 0.163 & 0.097 \\ 
\hline\hline
\end{tabular}
\label{table:EWRGBBempirical} 
\end{table*}

\section{Models}
\label{sec:Models}
We primarily use isochrones downloaded from the BaSTI stellar database\footnote{http://basti.oa-teramo.inaf.it}  \citep{2004ApJ...612..168P,2007AJ....133..468C}. Our tests span a broad range of input physics, including those that are $\alpha$-enhanced \citep{2004ApJ...616..498C,2006ApJ...642..797P}, modified helium abundances, modified ages, and that have have modified CNONa mixtures \citep{2009ApJ...697..275P} as per recent observations of Galactic GCs.  The CNONa enhanced mixture changing the effective mass ratios of C:N:O:Na:Fe from 0.11:0.035:1:0.0013:0.032 to 0.18:14:1:0.052:0.30, in other words, significant increases in carbon, nitrogen, and sodium relative to oxygen, with the fraction of the total mass of metals locked up in the elements CNONa rising from 88\% to 93\%.  We also look at some extreme metallicities for the sake of completeness \citep{2013A&A...558A..46P}. 

Models from the Dartmouth stellar database \citep{2008ApJS..178...89D} are also used as a theoretical comparison over the full metallicity space of the Galactic GC system and for standard ages and chemical mixtures. Those predictions are listed in Table \ref{table:PredictedRGBBParametersDartmouth}. 

\begin{figure*}
\begin{center}
\includegraphics[totalheight=0.65\textheight]{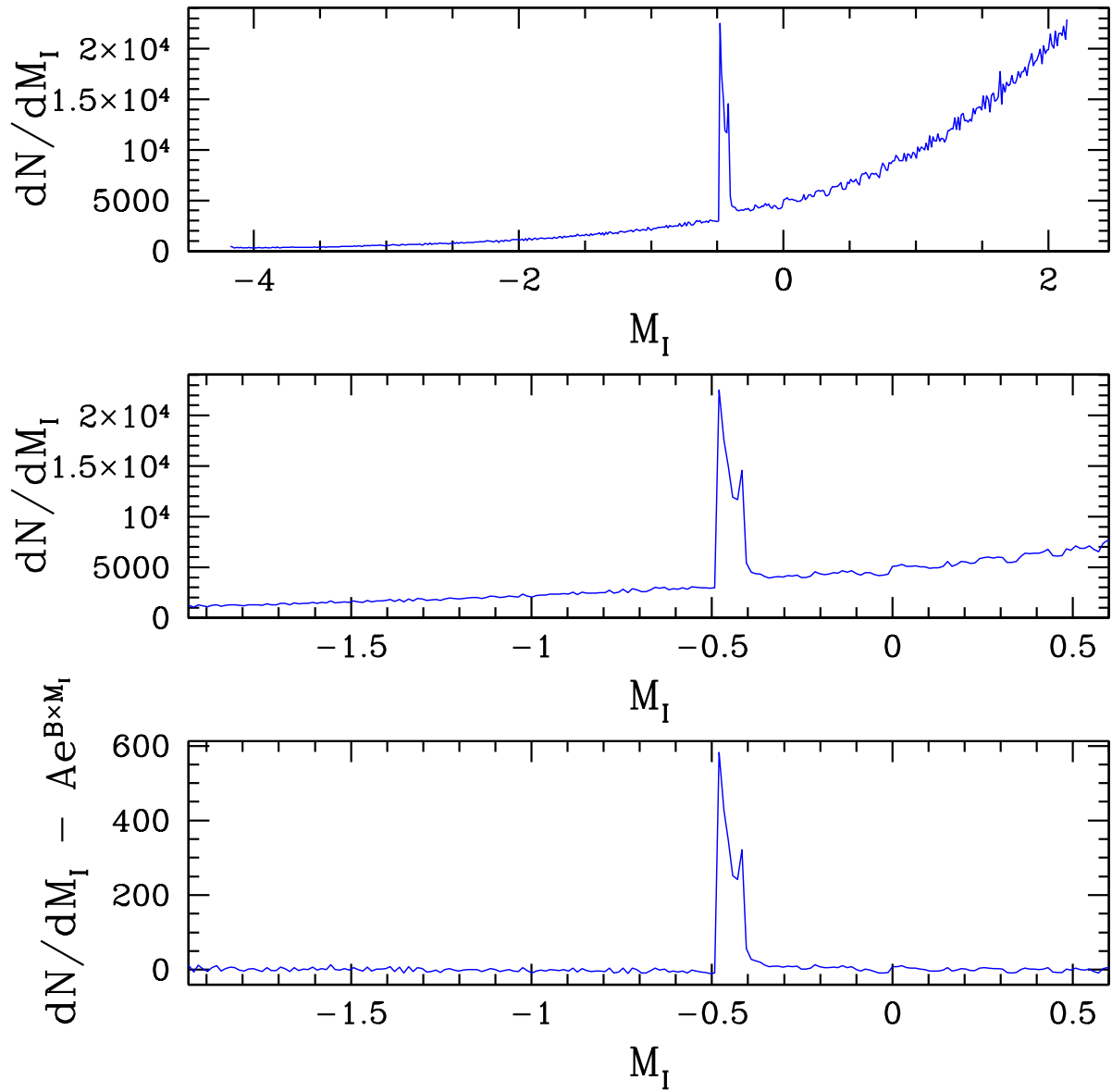}
\end{center}
\Large
\caption{\large  The BaSTI luminosity function \citep{2004ApJ...612..168P} for an [M/H]$=-0.963$, [$\alpha$/Fe]$=0.00$, $t=12 Gyr$ isochrone with standard helium and CNONa abundances.  This isochrone has a predicted exponential slope of $B=0.735$, $M_{I,RGBB}=-0.45$, and $EW_{\rm{RGBB}}=0.291$.  TOP: The luminosity function in the range $-4.18 \leq M_{I} \leq 2.14$,  over which the exponential form of the underlying red giant luminosity function away from the red giant branch bump is easily discernible. MIDDLE: The luminosity function over the range $I_{RGBB} - 1.50 \leq I \leq I_{RGBB} + 1.0$, which is where we compute our fits. BOTTOM: Residuals to the luminosity function with an exponential fit to the underlying red giant branch subtracted. The residuals are consistent with zero away from the RGBB.   The two peaks in the luminosity function correspond to the brightest and faintest parts of the RGBB, the reader may find a detailed discussion of the shape of the RGBB in \citet{2002ApJ...565.1231C}.  } 
\label{Fig:ModelLF}
\end{figure*}

For each isochrone, we sample the luminosity function (LF) in the magnitude range $I_{RGBB} - 1.50 \leq I \leq I_{RGBB} + 1.0$. We assume a Salpeter initial mass function \citep{1955ApJ...121..161S} such that $N(m) \propto m^{-2.35}$ when constructing the luminosity function. An exponential is fit to the luminosity function outside the RGBB, we then subtract this fit from the RGBB, measure the excess, and divide this excess by the best-fit number density of underlying RG stars at the luminosity of the RGBB. This yields a predicted value of $EW_{\rm{RGBB}}$ in a manner analogous to how astronomers routinely measure equivalent widths in stellar spectra. This process is further explained in Figure \ref{Fig:ModelLF}. Some related LF parameters for some of these isochrones were calculated and summarised by \citet{2014MNRAS.442.2075N}.

\subsection{The Effect of Varying the Total Metallicity [M/H]}
\begin{figure*}
\begin{center}
\includegraphics[totalheight=0.65\textheight]{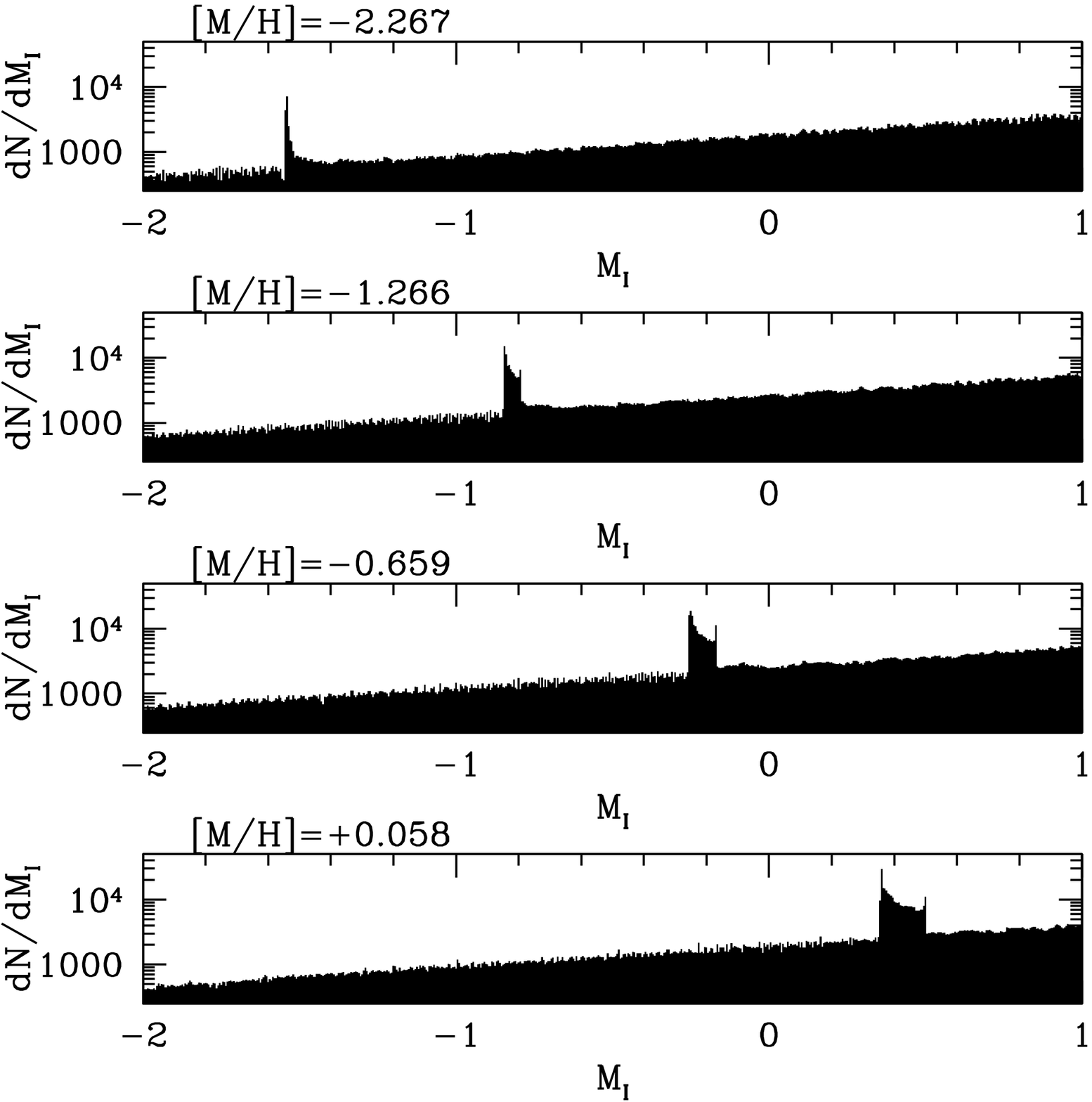}
\end{center}
\Large
\caption{\large  The BaSTI luminosity function \citep{2004ApJ...612..168P} for $t=12$ Gyr, [$\alpha$/Fe]$=+0.40$ for the red giant branch. As the metallicity varying from [M/H]$=-2.267$ in the top panel to  [M/H]$=+0.058$ in the bottom panel, the normalisation of the red giant branch bump  increases from  $EW_{RGBB}=0.146$ to $EW_{RGBB}=0.424$. This increased normalisation is in addition to the underlying normalisation of the red giant branch at the location of the bump itself increasing.  } 
\label{Fig:ModelLFmetallicityeffect}
\end{figure*}

What we have found is that the total metallicity [M/H] and the initial helium abundance $Y$ have a substantial effect on the predicted value of $EW_{\rm{RGBB}}$, whereas the age $t$/Gyr  and the details of the metallicity mixture ([$\alpha$/Fe] and CNONa variations at fixed total [M/H]) have minor effects. In Table \ref{table:PredictedRGBBParameters1}, we summarise the predicted values of $EW_{\rm{RGBB}}$ over a broad range of metallicities ($-2.27 \leq  \rm{[M/H]} \leq +0.40$) for standard $\alpha$-element, helium and CNONa abundances and an age  $t=12$ Gyr thought to be typical of Galactic GCs \citep{2009ApJ...694.1498M,2010ApJ...708..698D}, we also plot the predicted luminosity functions in Figure \ref{Fig:ModelLFmetallicityeffect}.  The expectation is from an increase of $EW_{\rm{RGBB}}=0.146$ mag at [M/H]$=-2.27$ to $EW_{\rm{RGBB}}=0.429$ mag at [M/H]$=+0.06$, with $EW_{\rm{RGBB}}$ then reaching a plateau maintained up to the highest metallicity probed [M/H]$=+$0.40. No simple predicted relationship can be written down for the dependence of $EW_{\rm{RGBB}}$ as it is clearly curvilinear, however, over the range $-1.79 \leq \rm{[M/H]} \leq -0.25$, it can be approximated by:
\begin{equation}
\begin{split}
EW_{\rm{RGBB}}  \approx 0.185 + 0.0732(\rm{[M/H]} + 1.79)+  \\ 
0.0411(\rm{[M/H]} + 1.79)^2 . 
\end{split}
\end{equation}

For real observations the detectability of the RGBB is expected to increase with metallicity at a rate higher than the factor of $0.429/0.146 \rightarrow 300\%$ one might naively infer from the numbers reported in the previous paragraph. In addition to this factor, the RGBB actually becomes fainter by $\sim 2$ mag and is thus shifted to a region of the RGB where stellar evolution is slower, and thus the underlying continuum against which we measure $EW_{\rm{RGBB}}$ is itself increasing. The actual increase in RGBB star counts between those two metallicities ([M/H]$=-$2.27,$+$0.06) should be of a factor of $\sim$10. 

We also looked at the model predictions from the Dartmouth stellar database \citet{2008ApJS..178...89D}, with the results listed in Table \ref{table:PredictedRGBBParametersDartmouth}. A comparison of the two curves can be found in the top panel of Figure \ref{Fig:EWRGBbumpVsMet}. The predictions from the Dartmouth models are systematically higher than those from the BaSTI models by $\sim$0.03 mag, with the offset being nearly consistent over the full range of metallicities. As BaSTI and Dartmouth assume similar values of the mixing length ($\alpha_{ML}=1.91,1.94$), the helium-enrichment ratio ($dY/dZ=1.4,1.5$) the primordial helium abundance ($Y_{BB}=0.245,0.245$), and the same opacities it is not clear why the predictions differ. \citet{2007IAUS..241...28W} noted that numerical issues may play a role, with the Dartmouth models predicting an RGBB luminosity $\sim$0.10 mag brighter than the FRANEC models (basis for the BaSTI database) even if all the parameters are fixed to be at the same value. 

\begin{table}
\caption{\large Predicted red giant branch bump equivalent width, $EW_{\rm{RGBB}}$ for standard Galactic globular cluster ages and chemistries over a broad range of metallicities, calculated from BaSTI isochrones \citep{2004ApJ...612..168P}, as a function of the metallicity [M/H], [$\alpha$/Fe] fixed to standard values, the age $t=12$ Gyr, and the initial helium abundance $Y$ fixed to scaled-solar, integrated in the luminosity range $I_{RGBB} - 1.50 \leq I \leq I_{RGBB} + 1.0$.  \newline}
\large
\centering 
\begin{tabular}{ccccc}
	\hline \hline
[M/H] & [$\alpha$/Fe] & $Y$ & t/Gyr &  $EW_{\rm{RGBB}}$  \\
	\hline \hline \hline
$-$2.27 & $+$0.40 & 0.245 & 12 & 0.146 \\ 
$-$1.79 & $+$0.40 & 0.245 & 12 & 0.185 \\ 
$-$1.49 & $+$0.40 & 0.246 & 12 & 0.212 \\ 
$-$1.27 & $+$0.40 & 0.246 & 12 & 0.232 \\ 
 $-$0.96 & $+$0.40 & 0.248 & 12 & 0.276 \\ 
$-$0.66 & $+$0.40 & 0.251 & 12 & 0.320 \\ 
$-$0.35 & $+$0.40 & 0.256 & 12 & 0.375 \\ 
$-$0.25 & $+$0.40 & 0.259 & 12 & 0.396 \\ 
$+$0.06 & 0.00 & 0.273 & 12 & 0.429 \\ 
$+$0.25 & 0.00 & 0.288 & 12 & 0.425 \\ 
$+$0.40 & 0.00 & 0.303 & 12 & 0.417 \\ 
 \hline
 \hline
\end{tabular}
\label{table:PredictedRGBBParameters1}
\end{table}

\begin{table}
\caption{\large Predicted red giant branch bump equivalent width, $EW_{\rm{RGBB}}$ for standard Galactic globular cluster ages and chemistries over a broad range of metallicities, calculated from Dartmouth isochrones \citep{2008ApJS..178...89D}. Symbols and methodology as in Table \ref{table:PredictedRGBBParameters1}. \newline}
\large
\centering 
\begin{tabular}{ccccc}
	\hline \hline
[M/H] & [$\alpha$/Fe] & $Y$ & t/Gyr &  $EW_{\rm{RGBB}}$  \\
	\hline \hline \hline
$-$2.20 & $+$0.40 & 0.245 & 12 & 0.168 \\ 
$-$1.70 & $+$0.40 & 0.246 & 12 & 0.228 \\ 
$-$1.20 & $+$0.40 & 0.247 & 12 & 0.269 \\ 
$-$0.70 & $+$0.40 & 0.251 & 12 & 0.344 \\ 
$-$0.20 & $+$0.40 & 0.262 & 12 & 0.438 \\ 
$+$0.07 & 0.00 & 0.274 & 12 & 0.461 \\ 
$+$0.21 & 0.00 & 0.286 & 12 & 0.462 \\ 
$+$0.36 & 0.00 & 0.301 & 12 & 0.459 \\ 
$+$0.56 & 0.00 & 0.329 & 12 & 0.429 \\ 
 \hline
 \hline
\end{tabular}
\label{table:PredictedRGBBParametersDartmouth}
\end{table}

\subsection{The Effect of Varying the Initial Helium Abundance $Y$}
The second most significant effect is that of variations in the initial helium abundance $Y$.  In Table \ref{table:PredictedRGBBParameters2}, we summarise the predicted effect on $EW_{\rm{RGBB}}$ of increasing $Y$ for four different metallicities. Increasing $Y$ is predicted to decrease the value of $EW_{\rm{RGBB}}$. At [M/H]$=-2.27$, the expectation is from an decrease of $EW_{\rm{RGBB}}=0.146$ at $Y=0.245$ to $EW_{\rm{RGBB}}=0.053$ at $Y=0.400$. The fractional decrease remains high though is a little slower at [M/H]$=+0.06$, where the expectation is from an decrease of $EW_{\rm{RGBB}}=0.429$ at $Y=0.273$ to $EW_{\rm{RGBB}}=0.220$ at $Y=0.400$.  The predicted relationship can be summarised as follows:
\begin{equation}
\frac{d\,EW_{\rm{RGBB}}}{dY}  \approx - [0.85 + 0.49(\rm{[M/H]} + 1.79)] . 
\end{equation}

It will be interesting to see if this prediction is verified in Galactic globular clusters as multi-colour photometry continues to refine the splits in red giant branches due to metallicity, CNONa abundances, helium, \textit{et cetera}. This decrease in $EW_{\rm{RGBB}}$ with increasing $Y$ is in addition to the fact that increasing $Y$ raises the luminosity of the RGBB ($dM_{I}/dY \sim -3.0$), shifting the RGBB to a region of slightly faster evolution, as well as the fact that the total normalisation of the RGB lifetime decreases with increasing $Y$ at a rate of ${\Delta}\rm{Log}\,t_{\rm{RGB}} \sim -0.84{\Delta}Y$ \citep{1994A&A...285L...5R}. The combination of these two factors should decrease the RGBB star counts by $\sim$35\% for a change of ${\Delta}Y=+0.10$, in addition to the effect of a decreased  $EW_{\rm{RGBB}}$. 

\begin{table}
\caption{\large Predicted red giant branch bump equivalent width,  $EW_{\rm{RGBB}}$ , for four metallicities, standard ages and metallicities, but over a range of initial helium abundances. Symbols and methodology as in Table \ref{table:PredictedRGBBParameters1}.  \newline}
\large
\centering 
\begin{tabular}{ccccc}
	\hline \hline
[M/H] & [$\alpha$/Fe] & $Y$ & t/Gyr &  $EW_{\rm{RGBB}}$  \\
	\hline \hline \hline
$-$1.79 & $+$0.40 & 0.245 & 12 & 0.185 \\ 
$-$1.79 & $+$0.40 & 0.300 & 12 & 0.094  \\ 
$-$1.79 & $+$0.40 & 0.350 & 12 & 0.080 \\ 
$-$1.79 & $+$0.40 & 0.400 & 12 &  0.053 \\ 
$-$0.96 & $+$0.40 & 0.248 & 12 & 0.276 \\
$-$0.98 & $+$0.40 & 0.300 & 12 & 0.201  \\ 
$-$0.96 & $+$0.40 & 0.350 & 12 &  0.147 \\ 
$-$0.96 & $+$0.40 & 0.400 & 12 &  0.100 \\ 
$-$0.35 & $+$0.40 & 0.256 & 12 & 0.375 \\ 
$-$0.35 & $+$0.40 & 0.300 & 12 & 0.296  \\ 
$-$0.35 & $+$0.40 & 0.350 & 12 &  0.237 \\ 
$-$0.35 & $+$0.40 & 0.350 & 12 &  0.159 \\ 
$+$0.06 & $+$0.40 & 0.273 & 12 & 0.424 \\ 
$+$0.07 & $+$0.40 & 0.350 & 12 &  0.289 \\ 
$+$0.06 & $+$0.40 & 0.400 & 12 &  0.220 \\ 
 \hline
 \hline
\end{tabular}
\label{table:PredictedRGBBParameters2}
\end{table}

\subsection{The Effect of Varying the Age at Fixed Metallicity}
The models predict that age variations at the level observed in Galactic GCs should yield a minor impact on measurements of $EW_{\rm{RGBB}}$. We show predicted $EW_{\rm{RGBB}}$ for four representative metallicities and ages $t=4,7,10,12,\,\rm{and}\,14$ Gyr in Table \ref{table:PredictedRGBBParameters4}. Decreasing age is predicted to increase $EW_{\rm{RGBB}}$ at low metallicity. At [M/H]$=-0.35$, this trend begins to reverse, with $EW_{\rm{RGBB}}$ decreasing between $t=7$Gyr and $t=4$Gyr, and for [M/H]$=+0.25$ the peak in $EW_{\rm{RGBB}}$ is reached in the age range $10 < t/\rm{Gyr} < 14$. 

For the age range $10 < t/\rm{Gyr} < 14$, which is the one relevant to Galactic GCs \citep{2009ApJ...694.1498M,2010ApJ...708..698D}, the predicted relationship can be summarised as follows:
\begin{equation}
\frac{d\,EW_{\rm{RGBB}}}{dt}  \approx -0.016 + 0.0065(\rm{[M/H]} + 2.27) \, . 
\end{equation}

\begin{table}
\caption{\large Predicted red giant branch bump equivalent width, for four metallicities but over a range of ages. Symbols and methodology as in Table \ref{table:PredictedRGBBParameters1}.  \newline}
\large
\centering 
\begin{tabular}{ccccc}
	\hline \hline
[M/H] & [$\alpha$/Fe] & $Y$ & t/Gyr &  $EW_{\rm{RGBB}}$  \\
	\hline \hline \hline
$-$1.79 & $+$0.40 & 0.245 & 14 & 0.137 \\ 
$-$1.79 & $+$0.40 & 0.245 & 12 & 0.185 \\ 
 $-$1.79 & $+$0.40 & 0.245 & 10 & 0.211 \\ 
 $-$1.79 & $+$0.40 & 0.245 & 7 & 0.260 \\ 
 $-$1.79 & $+$0.40 & 0.245 & 4 & 0.327 \\ 
 $-$0.96 & $+$0.40 & 0.248 & 14 & 0.259 \\ 
  $-$0.96 & $+$0.40 & 0.248 & 12 & 0.276 \\ 
 $-$0.96 & $+$0.40 & 0.248 & 10 & 0.314 \\ 
 $-$0.96 & $+$0.40 & 0.248 & 7 & 0.336 \\ 
  $-$0.96 & $+$0.40 & 0.248 & 4 & 0.374 \\ 
  $-$0.35 & $+$0.40 & 0.256 & 14 & 0.364 \\ 
  $-$0.35 & $+$0.40 & 0.256 & 12 & 0.375 \\ 
  $-$0.35 & $+$0.40 & 0.256 & 10 & 0.389 \\ 
  $-$0.35 & $+$0.40 & 0.256 & 7 & 0.401 \\ 
  $-$0.35 & $+$0.40 & 0.256 & 4 & 0.375 \\ 
   $+$0.25 & 0.00 & 0.288 & 14 & 0.423 \\ 
  $+$0.25 & 0.00 & 0.288 & 12 & 0.425 \\ 
   $+$0.25 & 0.00 & 0.288 & 10 & 0.421 \\ 
  $+$0.25 & 0.00 & 0.288 & 7 & 0.398 \\ 
 $+$0.25 & 0.00 & 0.288 & 4 & 0.307 \\ 
 \hline
 \hline
\end{tabular}
\label{table:PredictedRGBBParameters4}
\end{table}

\subsection{The Effect of Varying the Metals Mixture at Fixed Total [M/H]}
\label{subsec:MetalsMixture}
It is well documented that Galactic GCs host stars of varying chemical mixtures, both in terms of $\alpha$-elements \citep{2009A&A...505..139C} and CNONa mixtures \citep{1999AJ....118.1273I,2009ApJ...695L..62Y,2011ApJ...730L..16M}. Fixing the total value of [M/H] is to fix the number of metallic ions relative to hydrogen, it does not necessarily mean that stellar structure should stay the same in all cases (in fact it does not), as different metallic species can contribute differently to the opacity of star, as well as be thermonuclear catalysts. We thus investigate the effect of varying [$\alpha$/Fe] at fixed [M/H], summarised in Table \ref{table:PredictedRGBBParameters5}, and of varying CNONa at fixed [M/H], summarised in Table \ref{table:PredictedRGBBParameters6}. We find that for both cases $EW_{\rm{RGBB}}$ is predicted to be nearly independent of the metals mixture.

\begin{table}
\caption{\large Predicted red giant branch bump equivalent width as $\alpha$-abundances are decreased and Fe-abundances increased at fixed [M/H]. Symbols and methodology as in Table \ref{table:PredictedRGBBParameters1}. \newline}
\large
\centering 
\begin{tabular}{ccccc}
	\hline \hline
[M/H] & [$\alpha$/Fe] & $Y$ & t/Gyr &  $EW_{\rm{RGBB}}$  \\
	\hline \hline \hline
$-$1.79 & $+$0.40 & 0.245 & 12 & 0.185 \\ 
$-$1.79 & 0.00 & 0.245 & 12 & 0.188 \\ 
 $-$0.96 & $+$0.40 & 0.248 & 12 & 0.276 \\ 
  $-$0.96 & 0.00 & 0.248 & 12 & 0.291 \\ 
$-$0.35 & $+$0.40 & 0.256 & 12 & 0.375 \\ 
$-$0.35 & 0.00 & 0.256 & 12 & 0.382 \\ 
$+$0.25 &  $+$0.40  & 0.288 & 12 & 0.424 \\ 
$+$0.25 & 0.00 & 0.288 & 12 & 0.425 \\ 
 \hline
 \hline
\end{tabular}
\label{table:PredictedRGBBParameters5}
\end{table}

\begin{table*}
\caption{\large Predicted red giant branch bump equivalent width, for four different representative combinations of initial helium abundance and [M/H], but each with different combination having two different partitions between CNONa and Fe with the total metallicity fixed. Symbols and methodology mostly as in Table \ref{table:PredictedRGBBParameters1}, we also list [Fe/H] and CNONa status.  \newline}
\large
\centering 
\begin{tabular}{ccccccc}
	\hline \hline
[M/H] & [Fe/H] & [$\alpha$/Fe] & CNONa extreme & $Y$ & t/Gyr &  $EW_{\rm{RGBB}}$  \\
	\hline \hline \hline
$-$1.79 & $-$2.14 & $+$0.40 & No & 0.245 & 12 & 0.185  \\ 
$-$1.79 & $-$2.41 & $+$0.40 & Yes & 0.245 & 12 &  0.180  \\ 
$-$0.96 & $-$1.31 & $+$0.40 & No & 0.248 & 12 & 0.276 \\
$-$0.96 & $-$1.58 & $+$0.40 & Yes & 0.248 & 12 &  0.257 \\
$-$0.96 &$-$1.31 &  $+$0.40 & No & 0.350 &   12 &  0.147 \\ 
$-$0.96 &$-$1.58 &  $+$0.40 &Yes & 0.350 &   12 &  0.139 \\ 
$-$0.35 & $-$0.70 & $+$0.40 & No & 0.256 & 12 & 0.375 \\ 
$-$0.35 & $-$0.97 & $+$0.40 & Yes & 0.256 & 12 & 0.321  \\ 
 \hline
 \hline
\end{tabular}
\label{table:PredictedRGBBParameters6}
\end{table*}

A decrease in [$\alpha$/Fe] of 0.40 dex, coupled to an increase in [Fe/H] of $\sim$0.29 dex such that [M/H] is fixed, does increase $EW_{\rm{RGBB}}$, but barely so. The predicted shift ${\Delta}EW_{\rm{RGBB}}$ is never higher than $\sim$0.015 mag, and typically $\sim$0.005 mag. Similarly, increasing CNONa at fixed [M/H] such that [Fe/H] decreases by 0.27 dex does increase $EW_{\rm{RGBB}}$ by a small amount, though larger than for shifts due to $\alpha$-enhancement. The predicted shift is ${\Delta}EW_{\rm{RGBB}}=+0.018$ mag at [M/H]$=-1.79$ and  ${\Delta}EW_{\rm{RGBB}}=-0.036$ mag at [M/H]$=-0.35$. For both cases, the predicted shift is vastly smaller than the median measurement error of $\sim$0.08 mag reported by \citet{2013ApJ...766...77N}. The takeaway is that $EW_{\rm{RGBB}}$ will depend far more on the total metallicity than on the details of the metals mixture.

\section{$EW_{\rm{RGBB}}$ in Data Versus Models}
\begin{figure}
\begin{center}
\includegraphics[totalheight=0.39\textheight]{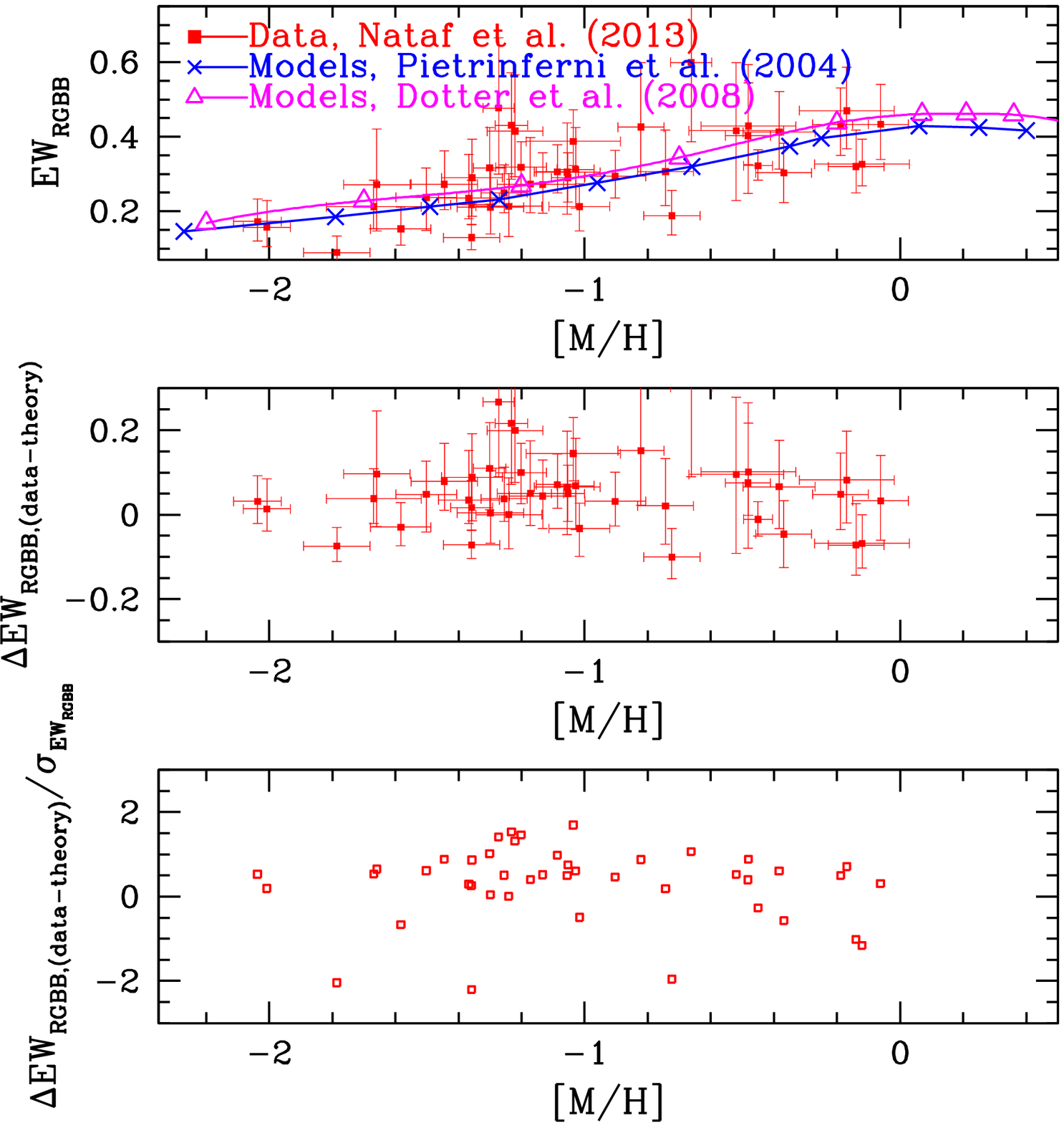}
\end{center}
\Large
\caption{\large  TOP: The trend of the equivalent width of the RGBB, $EW_{\rm{RGBB}}$, versus metallicity. Empirical values (shaded red squares) are from \citet{2013ApJ...766...77N}, and the theoretical values under the assumption of $t=12$ Gyr and a standard elemental mixture are taken from  the BaSTI stellar database (blue X's, \citealt{2004ApJ...612..168P}) and the Dartmouth stellar database (unshaded magenta triangles,  \citealt{2008ApJS..178...89D}). Model predictions are interpolated by cubic spline. MIDDLE: Residuals when the predicted $EW_{\rm{RGBB}}$ are subtracted from the observed values.  BOTTOM: Residuals from the middle panel, with their values divided by the error taking the asymmetry of the error bars into account. } 
\label{Fig:EWRGBbumpVsMet}
\end{figure}

In Figure \ref{Fig:EWRGBbumpVsMet} we plot the measured and predicted $EW_{\rm{RGBB}}$ in the top panel, with the residuals relative to the BaSTI models obtained when one subtracts the predicted from the measured values in the middle and bottom panels, all as a function of [M/H]. 

The weighted-linear least squares fit yields an offset relative to the BaSTI models of: 
\begin{equation}
\begin{split}
EW_{\rm{RGBB,(data-predicted)}} = (-0.014 \pm 0.011) + \\ 
(0.002 \pm 0.021)(\rm{[M/H]} + 1.0216) ),
\end{split}
\end{equation}
and relative to the Dartmouth models of:
\begin{equation}
\begin{split}
EW_{\rm{RGBB,(data-predicted)}} = (-0.045 \pm 0.011) + \\ 
(0.001 \pm 0.021)(\rm{[M/H]} + 1.0216) ),
\end{split}
\end{equation}
in other words, the observed $EW_{\rm{RGBB}}$ are smaller than the predicted values, by a  small $(0.014 \pm 0.011)$ mag in the BaSTI models and by $(0.045 \pm 0.011)$ in the Dartmouth models. The trend with metallicity is consistent with zero for both sets of models.


These results undermine the possibility that the offset between predicted and observed  RGBB luminosities is due to a calibration error in the metallicity scale of Galactic GCs.  That problem, viewed in isolation, could be resolved if the metallicity scale of Galactic GCs \citep{2009A&A...508..695C} \textit{underestimated} metallicities. However, an analogously perfect match RGBB star count predictions and observations would require that the metallicity scale \textit{overestimate} metallicities. The offsets relative to the BaSTI and Dartmouth predictions would respectively be brought to zero if the Galactic GC metallicity scale is assumed to be overestimated by a linear shift of $\sim 0.11$ dex and $\sim 0.36$ dex. A shift in the metallicity scale is thus not a viable proposition for the data and model offset in RGBB physics.

\subsection{A Note on Comparing $EW_{\rm{RGBB}}$ in Data and Models In Light of the Error in Predicted RGBB Luminosities}
\label{subsec:cancellation}
The naive ``null-hypothesis" one should first attempt is a straight-up comparison between the predicted and observed values of  $EW_{\rm{RGBB}}$, However, as noted in the introduction, it is by now well-documented that theory overestimates the luminosity of the RGBB by a typical value of $\sim$0.20 mag, reaching $\sim$0.40 mag in the most metal-poor clusters \citep{2010ApJ...712..527D,2011A&A...527A..59C,2011PASP..123..879T}. If one were to then assume that what theory should predict is the \textit{integrated luminosity} of the RGBB, one might then shift the comparison point. 

A coincidental cancellation ends up mitigating this concern. As the RGBB is observed to take place at a magnitude $\sim$0.20 mag dimmer than predicted by models, the assumption of fixed integrated luminosity would yield a duration for this phase of stellar evolution extended by $\sim$17\% relative to model predictions. However, the RG branch against which $EW_{\rm{RGBB}}$ is normalised is denser at these magnitudes, by an approximate factor exp($0.73 \times 0.20) = 1.16$, or 16\%, which nearly perfectly offsets the aforementioned 17\%. The respective numbers are 31\% and 34\% if one assumes a 0.40 magnitude offset, once again similar.  

This cancellation is due to the fact that the exponential component of the RG luminosity function goes as $N(m) \propto \exp(0.73 \times M_{\rm{Bol}}) \approx 10^{(0.32 \times M_{\rm{Bol}})}$ \citep{1989A&A...216...62C,2013ApJ...766...77N}, whereas luminosities goes as $L \propto 10^{-0.40 \times M_{\rm{Bol}}}$. These two effects combine for a distortion of $\sim 10^{0.08 \times {\Delta} M_{\rm{Bol}}}$, where ${\Delta}M_{\rm{Bol}}$ is the offset between predicted and observed magnitudes, which is smaller than 10\% for offsets of ${\Delta}M \leq 0.50$ mag.  

\subsection{A Note on Comparing $EW_{\rm{RGBB}}$ in Data and Models In Light of Multiple Stellar Populations}
\label{subsec:Multiple}
The anonymous referee expressed concern about the possible role of multiple stellar populations in affecting the results and interpretations here. We have excluded 4 GCs from our analysis (NGC 2808, 5286, 6388, and 6441) but there is a possibility that other GCs might have just as much internal diversity. 

We recognise this concern, and that it is possible that in the future there will be a need to re-visit this analysis for these reasons. As analytical methods improve, more GCs might turn out to be more challenging to interpret. \citet{2013MNRAS.434.3542Y} recently measured an intrinsic spread of ${\sigma}$[Fe/H] $\approx 0.03$ dex for NGC 6752. That is too small to affect interpretations of the RGBB, but in principal the same methodology could be applied to other clusters leading to larger measured spreads that would matter. Indeed, \citet{2014MNRAS.441.3396Y} measured a much larger variation for NGC 7089 (M2). They found that 4 of 14 stars (29\%) from their biased sample had their metallicity enhanced from [Fe/H]$ \approx -$1.7 (the mode of the cluster metallicity distribution function) to [Fe/H]$ \approx -1.0$. This would in principle explain the underpopulated RGB bump of NGC 7089 as a number of the stars would be undergoing the RGBB phase at a different luminosity. However, \citet{2012ApJ...760...39P} showed, using a photometric analysis of the subgiant branch, that only $\sim$5\% of the stars in that cluster belonged to the extreme population, which is too small to noticeably affect our results. Indeed, $EW_{RGBB,\rm{NGC\,7089}}=0.129 \pm 0.033$, whereas the predicted value for [M/H]$=-1.36$ is $EW_{RGBB}=0.223$ for the BaSTI models and $EW_{RGBB}=0.254$ for the Dartmouth models. The offset is closer to 50\% than it is to 5\%, thus there may be another factor at play, or simply a statistical fluke. The globular clusters NGC 5024 (M53) and NGC 6723 show comparable deficiencies in their RGBB star counts to NGC 7089 (M2), and thus may warrant precision abundance investigations. 

We showed in Section \ref{subsec:MetalsMixture} that varying the metals mixture at fixed total metallicity has little impact on $EW_{RGBB}$. With that said, one should maintain awareness of these variations when measuring RGBB parameters. First of all, CNONa does not vary at fixed [M/H], but rather at fixed [Fe/H], as such the CNONa-enhanced population will be [M/H]-enhanced and will  thus be expected to have more densely-populated RGBBs. Further, though $EW_{\rm{RGBB}}$ is predicted to be largely insensitive to the details of the metals mixture, $M_{I,RGBB}$ is not. Decreasing [$\alpha$/Fe] by 0.40 dex at fixed [M/H] dims the RGBB by a measurable ${\Delta}M_{I,RGBB} \sim$0.07 mag, whereas increasing CNONa brightens the RGBB by $\sim$0.03 mag at [M/H]$=-$1.79,  and by a much larger $\sim$0.17 mag at [M/H]$=-$0.35.  These shifts are sufficiently large that they can either distort the shape of the RGBB or even split it into separate RGBBs localised at separate luminosities, necessitating care in analysis. Once again, as more data emerges on the multiple populations in Galactic GCs, there may be a need for a re-analysis of RGBB physics in data and models. 



\section{Discussion and Conclusion}
\label{sec:Discussion}
We have compiled the best literature data available on $EW_{\rm{RGBB}}$ \citep{2013ApJ...766...77N}, a parameter of star counts on the RGBB, and compared these to predictions from the BaSTI stellar database \citep{2004ApJ...612..168P}. We have found that the offset between data and theory is small, $EW_{\rm{RGBB,(data-predicted)}} = (0.014 \pm 0.011)$, consistent with zero for the BaSTI models, with the offset rising to  $EW_{\rm{RGBB,(data-predicted)}} = (0.045 \pm 0.011)$ for the Dartmouth models.

These results constrain suggested solutions to the issue of the offset between predicted and observed RGBB luminosities in Galactic GCs \citep{2010ApJ...712..527D,2011A&A...527A..59C,2011PASP..123..879T}. That problem cannot be resolved by assuming an error in the adopted metallicity scale, as a systematic shift as small as 0.16 dex in [M/H] (not sufficient to resolve the luminosity discrepancy) would yield a $\geq$3$\sigma$ offset between measured and predicted $EW_{\rm{RGBB}}$. The issue is thus more likely to be one pertaining to the \textit{physics} of stellar models, such as the treatment of convection, as suggested by  \citet{1991A&A...244...95A} and \citet{2012ApJ...746...20K}.  

Looking to the horizon, we see the prospects for improved data from Galactic GCs as marginal, as the existing surveys are already accomplished \citep{2002A&A...391..945P,2007AJ....133.1658S,2011ApJ...738...74D}. There may be opportunities for improved measurements of Galactic bulge GCs as multi-colour photometry comes in allowing a treatment of the substantial differential reddening observed toward the inner Galaxy, thus constraining the behaviour of the RGBB at higher metallicities. We argue that this consistency increases confidence in RGBB star counts as an actionable parameter with which to constrain the nature of stellar populations, see also \citet{2011ApJ...730..118N} and \citet{2011ApJ...736...94N}. The theoretical predictions listed within this work extend beyond the age-helium-metallicity space spanned by the Galactic globular cluster system, and thus can be of use to other kinds of observation, for example that of luminosity functions investigated toward the \textit{Kepler} field \citep{2011A&A...530A.100H}.

\section*{Acknowledgments}
We thank Martin Asplund, Santi Cassisi, and Aaron Dotter for helpful discussions. We thank the anonymous referee for a helpful report.
DMN was  supported by the Australian Research Council grant FL110100012. 
This work has made use of BaSTI web tools.


\begin{thebibliography}{42}
\providecommand{\natexlab}[1]{#1}
\bibitem[{{Alongi} et~al.(1991){Alongi}, {Bertelli}, {Bressan} \&
  {Chiosi}}]{1991A&A...244...95A}
{Alongi} M., {Bertelli} G., {Bressan} A., {Chiosi} C., 1991, \aap, 244, 95

\bibitem[{{Bjork} \& {Chaboyer}(2006)}]{2006ApJ...641.1102B}
{Bjork} S.~R., {Chaboyer} B., 2006, \apj, 641, 1102

\bibitem[{{Bonatto} \& {Bica}(2008)}]{2008A&A...479..741B}
{Bonatto} C., {Bica} E., 2008, \aap, 479, 741

\bibitem[{{Bono} et~al.(2001){Bono}, {Cassisi}, {Zoccali} \&
  {Piotto}}]{2001ApJ...546L.109B}
{Bono} G., {Cassisi} S., {Zoccali} M., {Piotto} G., 2001, \apjl, 546, L109

\bibitem[{{Carretta} et~al.(2009{\natexlab{a}}){Carretta}, {Bragaglia},
  {Gratton}, {D'Orazi} \& {Lucatello}}]{2009A&A...508..695C}
{Carretta} E., {Bragaglia} A., {Gratton} R., {D'Orazi} V., {Lucatello} S.,
  2009{\natexlab{a}}, \aap, 508, 695

\bibitem[{{Carretta} et~al.(2009{\natexlab{b}}){Carretta}, {Bragaglia},
  {Gratton} \& {Lucatello}}]{2009A&A...505..139C}
{Carretta} E., {Bragaglia} A., {Gratton} R., {Lucatello} S.,
  2009{\natexlab{b}}, \aap, 505, 139

\bibitem[{{Cassisi} \& {Salaris}(1997)}]{1997MNRAS.285..593C}
{Cassisi} S., {Salaris} M., 1997, \mnras, 285, 593

\bibitem[{{Cassisi} et~al.(2002){Cassisi}, {Salaris} \&
  {Bono}}]{2002ApJ...565.1231C}
{Cassisi} S., {Salaris} M., {Bono} G., 2002, \apj, 565, 1231

\bibitem[{{Cassisi} et~al.(2004){Cassisi}, {Salaris}, {Castelli} \&
  {Pietrinferni}}]{2004ApJ...616..498C}
{Cassisi} S., {Salaris} M., {Castelli} F., {Pietrinferni} A., 2004, \apj, 616,
  498

\bibitem[{{Cassisi} et~al.(2011){Cassisi}, {Mar{\'{\i}}n-Franch}, {Salaris},
  {Aparicio}, {Monelli} \& {Pietrinferni}}]{2011A&A...527A..59C}
{Cassisi} S., {Mar{\'{\i}}n-Franch} A., {Salaris} M., {Aparicio} A., {Monelli}
  M., {Pietrinferni} A., 2011, \aap, 527, A59

\bibitem[{{Castellani} et~al.(1989){Castellani}, {Chieffi} \&
  {Norci}}]{1989A&A...216...62C}
{Castellani} V., {Chieffi} A., {Norci} L., 1989, \aap, 216, 62

\bibitem[{{Cordier} et~al.(2007){Cordier}, {Pietrinferni}, {Cassisi} \&
  {Salaris}}]{2007AJ....133..468C}
{Cordier} D., {Pietrinferni} A., {Cassisi} S., {Salaris} M., 2007, \aj, 133,
  468

\bibitem[{{Di Cecco} et~al.(2010)}]{2010ApJ...712..527D}
{Di Cecco} A. et~al., 2010, \apj, 712, 527

\bibitem[{{Dotter} et~al.(2008){Dotter}, {Chaboyer}, {Jevremovi{\'c}},
  {Kostov}, {Baron} \& {Ferguson}}]{2008ApJS..178...89D}
{Dotter} A., {Chaboyer} B., {Jevremovi{\'c}} D., {Kostov} V., {Baron} E.,
  {Ferguson} J.~W., 2008, \apjs, 178, 89

\bibitem[{{Dotter} et~al.(2011){Dotter}, {Sarajedini} \&
  {Anderson}}]{2011ApJ...738...74D}
{Dotter} A., {Sarajedini} A., {Anderson} J., 2011, \apj, 738, 74

\bibitem[{{Dotter} et~al.(2010)}]{2010ApJ...708..698D}
{Dotter} A. et~al., 2010, \apj, 708, 698

\bibitem[{{Fusi Pecci} et~al.(1990){Fusi Pecci}, {Ferraro}, {Crocker}, {Rood}
  \& {Buonanno}}]{1990A&A...238...95F}
{Fusi Pecci} F., {Ferraro} F.~R., {Crocker} D.~A., {Rood} R.~T., {Buonanno} R.,
  1990, \aap, 238, 95

\bibitem[{{Hekker} et~al.(2011)}]{2011A&A...530A.100H}
{Hekker} S. et~al., 2011, \aap, 530, A100

\bibitem[{{Iben}(1968)}]{1968Natur.220..143I}
{Iben} I., 1968, \nat, 220, 143

\bibitem[{{Ivans} et~al.(1999){Ivans}, {Sneden}, {Kraft}, {Suntzeff}, {Smith},
  {Langer} \& {Fulbright}}]{1999AJ....118.1273I}
{Ivans} I.~I., {Sneden} C., {Kraft} R.~P., {Suntzeff} N.~B., {Smith} V.~V.,
  {Langer} G.~E., {Fulbright} J.~P., 1999, \aj, 118, 1273

\bibitem[{{Kamath} et~al.(2012){Kamath}, {Karakas} \&
  {Wood}}]{2012ApJ...746...20K}
{Kamath} D., {Karakas} A.~I., {Wood} P.~R., 2012, \apj, 746, 20

\bibitem[{{King} et~al.(1985){King}, {Da Costa} \&
  {Demarque}}]{1985ApJ...299..674K}
{King} C.~R., {Da Costa} G.~S., {Demarque} P., 1985, \apj, 299, 674

\bibitem[Milone et al.(2012)]{2012ApJ...744...58M} Milone, A.~P., Piotto, 
G., Bedin, L.~R., et al.\ 2012, \apj, 744, 58 

\bibitem[{{Mar{\'{\i}}n-Franch} et~al.(2009)}]{2009ApJ...694.1498M}
{Mar{\'{\i}}n-Franch} A. et~al., 2009, \apj, 694, 1498

\bibitem[{{Marino} et~al.(2011){Marino}, {Villanova}, {Milone}, {Piotto},
  {Lind}, {Geisler} \& {Stetson}}]{2011ApJ...730L..16M}
{Marino} A.~F., {Villanova} S., {Milone} A.~P., {Piotto} G., {Lind} K.,
  {Geisler} D., {Stetson} P.~B., 2011, \apjl, 730, L16

\bibitem[{{Monelli} et~al.(2010){Monelli}, {Cassisi}, {Bernard}, {Hidalgo},
  {Aparicio}, {Gallart} \& {Skillman}}]{2010ApJ...718..707M}
{Monelli} M., {Cassisi} S., {Bernard} E.~J., {Hidalgo} S.~L., {Aparicio} A.,
  {Gallart} C., {Skillman} E.~D., 2010, \apj, 718, 707

\bibitem[{{Nataf} et~al.(2011{\natexlab{a}}){Nataf}, {Gould}, {Pinsonneault} \&
  {Stetson}}]{2011ApJ...736...94N}
{Nataf} D.~M., {Gould} A., {Pinsonneault} M.~H., {Stetson} P.~B.,
  2011{\natexlab{a}}, \apj, 736, 94

\bibitem[{{Nataf} et~al.(2011{\natexlab{b}}){Nataf}, {Udalski}, {Gould} \&
  {Pinsonneault}}]{2011ApJ...730..118N}
{Nataf} D.~M., {Udalski} A., {Gould} A., {Pinsonneault} M.~H.,
  2011{\natexlab{b}}, \apj, 730, 118

\bibitem[{{Nataf} et~al.(2013){Nataf}, {Gould}, {Pinsonneault} \&
  {Udalski}}]{2013ApJ...766...77N}
{Nataf} D.~M., {Gould} A.~P., {Pinsonneault} M.~H., {Udalski} A., 2013, \apj,
  766, 77

\bibitem[{{Nataf} et~al.(2014){Nataf}, {Cassisi} \&
  {Athanassoula}}]{2014MNRAS.442.2075N}
{Nataf} D.~M., {Cassisi} S., {Athanassoula} E., 2014, \mnras, 442, 2075

\bibitem[Norris(2004)]{2004ApJ...612L..25N} Norris, J.~E.\ 2004, \apjl, 
612, L25 

\bibitem[{{Pietrinferni} et~al.(2004){Pietrinferni}, {Cassisi}, {Salaris} \&
  {Castelli}}]{2004ApJ...612..168P}
{Pietrinferni} A., {Cassisi} S., {Salaris} M., {Castelli} F., 2004, \apj, 612,
  168

\bibitem[{{Pietrinferni} et~al.(2006){Pietrinferni}, {Cassisi}, {Salaris} \&
  {Castelli}}]{2006ApJ...642..797P}
{Pietrinferni} A., {Cassisi} S., {Salaris} M., {Castelli} F., 2006, \apj, 642,
  797

\bibitem[{{Pietrinferni} et~al.(2009){Pietrinferni}, {Cassisi}, {Salaris},
  {Percival} \& {Ferguson}}]{2009ApJ...697..275P}
{Pietrinferni} A., {Cassisi} S., {Salaris} M., {Percival} S., {Ferguson} J.~W.,
  2009, \apj, 697, 275

\bibitem[{{Pietrinferni} et~al.(2013){Pietrinferni}, {Cassisi}, {Salaris} \&
  {Hidalgo}}]{2013A&A...558A..46P}
{Pietrinferni} A., {Cassisi} S., {Salaris} M., {Hidalgo} S., 2013, \aap, 558,
  A46

\bibitem[{{Piotto} et~al.(2002)}]{2002A&A...391..945P}
{Piotto} G. et~al., 2002, \aap, 391, 945

\bibitem[Piotto et al.(2007)]{2007ApJ...661L..53P} Piotto, G., Bedin, 
L.~R., Anderson, J., et al.\ 2007, \apjl, 661, L53 

\bibitem[Piotto et al.(2012)]{2012ApJ...760...39P} Piotto, G., Milone, 
A.~P., Anderson, J., et al.\ 2012, \apj, 760, 39 

\bibitem[{{Renzini}(1994)}]{1994A&A...285L...5R}
{Renzini} A., 1994, \aap, 285, L5

\bibitem[{{Riello} et~al.(2003)}]{2003A&A...410..553R}
{Riello} M. et~al., 2003, \aap, 410, 553

\bibitem[{{Salpeter}(1955)}]{1955ApJ...121..161S}
{Salpeter} E.~E., 1955, \apj, 121, 161

\bibitem[{{Sarajedini} et~al.(2007)}]{2007AJ....133.1658S}
{Sarajedini} A. et~al., 2007, \aj, 133, 1658

\bibitem[{{Sweigart} et~al.(1990){Sweigart}, {Greggio} \&
  {Renzini}}]{1990ApJ...364..527S}
{Sweigart} A.~V., {Greggio} L., {Renzini} A., 1990, \apj, 364, 527

\bibitem[{{Thomas}(1967)}]{1967ZA.....67..420T}
{Thomas} H.~C., 1967, \zap, 67, 420

\bibitem[{{Troisi} et~al.(2011)}]{2011PASP..123..879T}
{Troisi} F. et~al., 2011, \pasp, 123, 879

\bibitem[Weiss et al.(2007)]{2007IAUS..241...28W} Weiss, A., Cassisi, S., 
Dotter, A., Han, Z., \& Lebreton, Y.\ 2007, IAU Symposium, 241, 28 

\bibitem[{{Yong} et~al.(2009){Yong}, {Grundahl}, {D'Antona}, {Karakas},
  {Lattanzio} \& {Norris}}]{2009ApJ...695L..62Y}
{Yong} D., {Grundahl} F., {D'Antona} F., {Karakas} A.~I., {Lattanzio} J.~C.,
  {Norris} J.~E., 2009, \apjl, 695, L62
  
  \bibitem[Yong et al.(2013)]{2013MNRAS.434.3542Y} Yong, D., Mel{\'e}ndez, 
J., Grundahl, F., et al.\ 2013, \mnras, 434, 3542 


\bibitem[Yong et al.(2014)]{2014MNRAS.441.3396Y} Yong, D., Roederer, I.~U., 
Grundahl, F., et al.\ 2014, \mnras, 441, 3396 

\end{thebibliography}

\end{document}